\documentclass[twocolumn,prb,aps,floats,floatfix,showpacs,amsmath,amssymb]{revtex4}
\usepackage{graphicx}
\usepackage{bm}
\usepackage{times}

\begin{document}
%
%
\newcommand{\cp}[1]{d_{#1}^\dagger}
\newcommand{\cm}[1]{d_{#1}}
\newcommand{\fp}[1]{f_{#1}^\dagger}
\newcommand{\fm}[1]{f_{#1}}
\newcommand{\dpp}[1]{c_{#1}^\dagger}
\newcommand{\dm}[1]{c_{#1}}
\newcommand{\e}{\varepsilon}
\newcommand{\s}{\sigma}
\newcommand{\w}{\omega}
\newcommand{\up}{\uparrow}
\newcommand{\down}{\downarrow}
\newcommand{\I}{\Im m}
\newcommand{\Tr}{\mathrm{Tr}}
%
%
\title{The Mesoscopic Kondo Box: A Mean-Field Approach}

\author{Rainer Bedrich$^1$, S\'ebastien Burdin$^{2,3}$, Martina Hentschel$^1$}
\affiliation{$^1$ Max Planck Institut f\"ur Physik komplexer Systeme, N\"othnitzer Str. 38, 01187 Dresden, Germany}
\affiliation{$^2$ Institut f\"ur Theoretische Physik, Universit\"at zu K\"oln, Z\"ulpicher Str. 77, 50937 K\"oln, Germany}
\affiliation{$^3$ Condensed Matter Theory Group, CPMOH, UMR 5798, Universit\'e de Bordeaux I, 
33405 Talence, France}

\date{\today}

%
%
\begin{abstract}
{We study the mesoscopic Kondo box, consisting of a quantum spin $1/2$ 
interacting with a chaotic electronic bath as can be realized by a 
magnetic impurity coupled to electrons on a quantum dot, 
using a mean-field approach for the Kondo interaction. Its numerical efficiency allows us to analyze 
the Kondo temperature, the local magnetic susceptibility, and the
conductance statistics for a large number of samples with 
energy levels obtained by random matrix theory. 
We see pronounced parity effects in the average values and in the probability distributions, depending on an even and odd electronic occupation of the
quantum dot, respectively. 
These parity effects are directly accessible in experiments.}
\end{abstract}
\maketitle

\section{Introduction}
Many-body effects have been a key interest in condensed matter physics
for many decades. A prime example is the Kondo effect~\cite{hewson} 
that, in its original context, refers to an increase of the resistance
with decreasing temperature below the characteristic Kondo 
temperature $T_K$ in metals containing magnetic impurities. The
significant progress in the fabrication of mesoscopic and nanoscopic systems 
has lead to many alternative realizations of Kondo 
systems.~\cite{goldhaber98,cronenwett98,nygard00,Delattre}
A nice example is the so-called Kondo box,~\cite{thimm99} where a
finite number of electrons, confined in a quantum dot, is coupled to a 
single magnetic impurity. 
Their discrete energy spectrum introduces the mean level spacing $\Delta$ as
new energy scale.
It has been shown that physical quantities strongly deviate from the 
metallic behavior for 
$\Delta \gtrsim T_K$, i.e., when the size of the Kondo cloud screening
the impurity would become larger than the size of the 
system.~\cite{SimonAffleck1, SimonAffleck2, thimm99, italians, balseiro}
For example, they show parity
effects, i.e., characteristic differences for an even and odd number
of electrons in the system.~\cite{nakamura99,park03,montambaux98}
Kondo physics in the presence of a chaotic dot geometry and disorder, 
respectively, has been studied within the 
Kondo disorder model (KDM)~\cite{miranda, mucciolo, cornaglia06} 
using Anderson's poor man's scaling approach~\cite{poorman} to
calculate the Kondo temperature.
However, poor man's scaling cannot describe the strong coupling phase 
below $T_{K}$, and non-perturbative methods are needed, for example 
quantum Monte Carlo simulations (QMC),~\cite{QMC} numerical
renormalization group~\cite{NRG, NRG2} or the non-crossing 
approximation.~\cite{NCA} 
These methods are either numerically expensive or not reliable for 
low temperatures $T\ll T_K$.~\cite{bickers} The QMC has been used by Kaul \textit{et al.}~\cite{baranger} to calculate the local magnetic susceptibility 
$\chi_0$ for three different realizations of a chaotic system showing 
that mesoscopic fluctuations lead to significant deviations from bulk 
universality. However, parity effects and systematic studies of mesoscopic fluctuations were neither considered here nor 
in the KDM studies.

In the present Letter we use a mean-field approach which had been 
developed initially for the single impurity Kondo 
model in a macroscopic metal,~\cite{yoshimori70} and later adapted 
to some particular mesoscopic Kondo systems.~\cite{SimonAffleck2, Aguado}
We first introduce the method in the framework of the chaotic Kondo
box model. 
We then discuss the resulting probability distributions
of $T_K$, $\chi_0$, and the conductance $G$. 
Their knowledge and the characteristic 
differences between even and odd electronic fillings make the
comparison with experiments a realistic endeavor for the near future.


\section{Model} 
We investigate the Kondo box model:~\cite{thimm99}
An electronic bath, e.g. quantum dot, with discrete energy levels $\e_l$, coupled to a local spin $\mathbf{S}=1/2$. 
The system is described by the Kondo Hamiltonian,~\cite{hewson}
\begin{equation}\label{eq:1}
H = \sum_{l\,\s} \left(\e_l - \mu\right) \cp{l\s}\cm{l\s} 
+ J_K \: \mathbf{S}\cdot\mathbf{s_0}\, .
\end{equation}
Operator $\cp{l\s}$ ($\cm{l\s}$) creates (annihilates) a dot electron 
with level index $l=1,\ldots,N$ and spin component $\s = \up,\down$. 
The chemical potential $\mu$, related to an external gate voltage, 
fixes the number of dot electrons $N_C$, which 
effectively accounts for Coulomb blockade. In addition, the case of a 
fixed $\mu=0$ rather than a fixed $N_C$ will be considered. 
The local spin density $\mathbf{s_0}$ of the dot electrons 
at the impurity position is given by 
\begin{align}
s_0^z &= \left(\cp{0\up}\cm{0\up}-\cp{0\down}\cm{0\down}\right)/2\, ,\\
s_0^+ &= \cp{0\up}\cm{0\down}\, , \\
s_0^- &= \cp{0\down}\cm{0\up}\, ,
\end{align}
where the local electron operator $\cm{0\s}$ is related to the dot 
electron operators by 
\begin{equation}
\cm{0\s}=\sum_l A_l^{*} \cm{l\s}\, .
\end{equation} 
The complex coefficients $A_l$ correspond to the non-interacting 
wave function amplitude of level $\e_l$ at the impurity site 
($\sum_l \left|A_l\right|^2 = 1$) and are assumed to be spin-independent.
The quantum dot is thus characterized by $N$
energy levels $\e_l$, distributed from $-W/2$ to $W/2$, 
and the corresponding amplitudes $A_l$. In this Letter, our focus will be on chaotic quantum 
dots where the $\e_l$ are realized within random matrix 
theory~\cite{metha} to be a Gaussian orthogonal ensemble 
unfolded to a constant density of states, and the intensities $\left|A_l\right|^2$,
obtained within the random wave model,~\cite{berry77} are Porter-Thomas
distributed. 
For reasons of comparison we also introduce a system with equidistant 
$\e_l$ and constant $A_l$ that we will refer to as the clean system, cf., e.g., Fig.~\ref{fig_TK}.

\section{Mean-field approximation}
For each given configuration, we treat the Kondo 
interaction in a mean-field approximation.~\cite{yoshimori70}
The magnetic impurity spin is thus described in terms of  
fermionic operators $f_{\up,\down}$ obeying the constraint 
$\fp{\up} \fm{\up} + \fp{\down} \fm{\down}=1$. The Kondo interaction
in~(\ref{eq:1}) then reads 
\begin{equation}
J_K \mathbf{S}\cdot\mathbf{s_0} = 
\frac{J_K}{2}\sum_{\s\,\s'} \fp{\s}\fm{\s'}\cp{0\s'}\cm{0\s} 
- \frac{J_K}{4}\sum_{\s} \cp{0\s}\cm{0\s}\, .
\end{equation} 
The first term describes the 
spin flip processes, while the second term corresponds to a local 
potential scattering and is not further considered in the following. 
The mean-field treatment of the Kondo box Hamiltonian invokes two 
approximations, 
(i) replacing the quartic spin flip term 
by a quadratic term, 
$r\sum_{\s} (\fp{\s}\cm{0\s}+ \mathrm{h.c.})$, where the effective 
hybridization $r$ is determined self-consistently by minimizing the
free energy, 
and (ii) introducing a static Lagrange multiplier $\lambda$, in order 
to fulfill (on average) the constraint of a single impurity spin. 
So the Hamiltonian (\ref{eq:1}) of the Kondo box in mean-field 
approximation reads $H=H_{\uparrow}+H_{\downarrow}$, with 
\begin{equation} \label{eq:6}
H_{\sigma} = \sum_l \left[
\left(\e_l-\mu\right)\cp{l\s}\cm{l\s}
+ r (A_l^{*}\fp{\s}\cm{l\s}+ \mathrm{h.c.})
\right] - \lambda \fp{\s}\fm{\s}  \, .
\end{equation}
Hereafter, since the mean-field approximation decouples the spin 
components, we suppress the spin index.
The mean-field parameter $r$, the Lagrange multiplier $\lambda$, and 
the chemical potential $\mu$, satisfy the self-consistency equations
\begin{eqnarray} 
\label{eq:7}
-\frac{r}{J_{K}} &=&
\left
\langle \fp{}\cm{0}\right\rangle
=
T \sum_{l} 
A_l^{*} \sum_{i\w_n} G_{\mathrm{lf}}\left(i\w_n\right) \:, \\ \label{eq:8}
\frac{1}{2} &=& \left\langle \fp{}\fm{} \right\rangle = T \sum_{i\w_n} G_{\mathrm{ff}}\left(i\w_n\right) \:, \\
\frac{N_C}{2}&=& \sum_{l} \left\langle \cp{l}\cm{l} \right\rangle =T \sum_{l} \sum_{i\w_n} G_{\mathrm{ll}}\left(i\w_n\right) \label{eq:Nc} \:, 
\end{eqnarray}
with the temperature 
$T$ and the fermionic Matsubara frequencies 
$\w_n = \left(2n+1\right)\pi T$. 
Here, the thermal average $\langle\cdots\rangle$ is expressed
in terms of one-particle Green's functions, which are calculated
from the mean-field Hamiltonian~(\ref{eq:6}) using equations of motion. We find 
\begin{align}
G_{\mathrm{ff}}\left(i\w_n\right) &= 
\left[i\w_n + \lambda - \left|r\right|^2 
\sum_l \frac{\left|A_l\right|^2}{i\w_n+\mu-\e_l} \right]^{-1}\:, 
\label{eqGff} \\
G_{\mathrm{lf}}\left(i\w_n\right) &= 
\frac{r A_l}{i\w_n+\mu-\e_l} G_{\mathrm{ff}}\left(i\w_n\right) \:, 
\label{eqGlf}  \\
G_{\mathrm{ll}}\left(i\w_n\right) &= 
\frac{1}{i\w_n+\mu-\e_l}\left[1 + 
\frac{\left|r\right|^2 \left|A_l\right|^2}{i\w_n+\mu-\e_l} 
G_{\mathrm{ff}}\left(i\w_n\right) \right] 
\label{eqGll}  \, .  
\end{align}

\section{Results}
We now present the results starting with a discussion
of $T_K$, followed by an analysis of $\chi_0$ and $G$. In contrast to 
previous studies of chaotic Kondo systems~\cite{baranger} we are able to distinguish 
between an even and odd number $N_C$ of dot electrons,
cf.~Eq.~(\ref{eq:Nc}). We find pronounced 
parity  effects in the probability distributions of $T_K$, $\chi_0$ 
and $G$. Solving only Eqs.~(\ref{eq:7})
and~(\ref{eq:8}), we can furthermore address the case of a non-fixed $N_C$, 
which corresponds to fixing $\mu=0$ as done by Kaul \textit{et al.}.\cite{baranger}
All the numerical results presented here are obtained for $N=100$
levels, and an electronic filling $N_C=101$ and $102$ in the odd and
even cases, respectively. 
We checked that $N$ is large enough for band edge effects to not play a role.

\begin{figure}
\includegraphics[width=8.6cm]{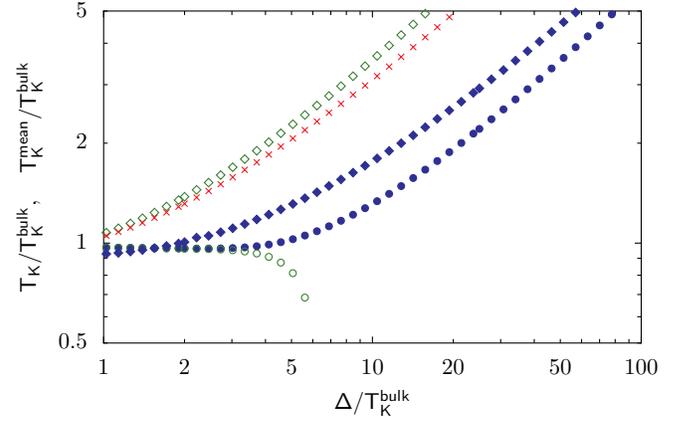}
\caption{(Color online) 
Kondo temperature $T_K$, for the clean system ($N_C$ even $\circ$, $N_C$ odd $\bullet$) and 
average Kondo temperature $T_K^{\mathrm{mean}}$, for the chaotic system 
($N_C$ even $\lozenge$, $N_C$ odd $\blacklozenge$, $\mu=0$ $\times$), 
as functions of $\Delta/T_K^{\mathrm{bulk}}$. 
\label{fig_TK}
 }
\end{figure}
\subsection{Kondo temperature}
Within the mean-field approximation, the Kondo temperature $T_{K}$ characterizes a transition between a
high temperature phase, where the effective hybridization $r$,
vanishes, and a low temperature phase with $r\neq 0$. More accurate
methods would rather describe a crossover, around $T_{K}$, 
between weakly and strongly coupled regimes. Despite the
oversimplified description of the $T>T_{K}$ weakly coupled regime, 
the mean-field 
approximation provides a good estimation for $T_{K}$, and a good
description of the physical properties in the $T<T_{K}$ strongly
coupled regime.~\cite{hewson} 
Using Eqs.~(\ref{eq:7}, \ref{eqGff}, \ref{eqGlf}), 
we find the well known Nagaoka-Suhl 
equation~\cite{nagaoka65} for $T_{K}$ 
\begin{equation}
\frac{2}{J_K} = \sum_l 
\frac{\left|A_l\right|^2}{\e_l-\mu} 
\tanh \left( \frac{\e_l-\mu}{2T_K}\right)\, .
\label{EqTK}
\end{equation}
In the bulk limit $\Delta / W \rightarrow 0$, one obtains the relation \cite{hewson}
\begin{equation}
T_K^{\mathrm{bulk}} \approx 1.13 
\sqrt{W^2 / 4-\mu^2} \, \exp\left(-W/J_K\right)\, .
\end{equation} 
In the following, we use $T_K^{\mathrm{bulk}}$, rather than $J_{K}$,
as the reference energy scale characterizing the 
Kondo coupling.

\begin{figure}
\includegraphics[width=8.6cm]{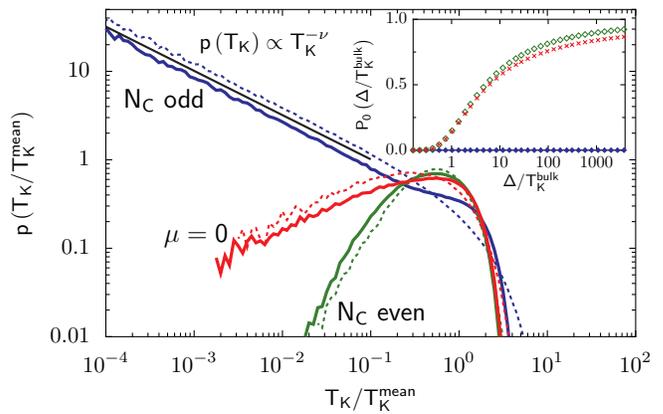}
\caption{(Color online) Kondo temperature distribution for a chaotic system with $\Delta/T_K^{\mathrm{bulk}}=2$ (solid lines), $100$ (dashed lines)
for fixed $N_C$ and $\mu=0$. For $N_C$ odd we see a power law $p(T_K)\propto T_{K}^{-\nu}$, indicated by the straight line. Inset: probability for vanishing $T_K$ in a chaotic system as function of $\Delta / T_K^{\mathrm{bulk}}$ for $N_C$ even ($\lozenge$), $N_C$ odd ($\blacklozenge$), and $\mu=0$ ($\times$).
 \label{fig_TKplaw}
 }
\end{figure}

The Kondo temperature probability distribution, $p(T_K)$, is studied by solving  
Eq.~(\ref{EqTK}) for various realizations. The result is mostly a regular, smooth function, which is depicted 
in the main part of Fig.~\ref{fig_TKplaw}. In addition to the data shown, there is a non-vanishing contribution $p(T_K=0)$ 
for $N_C$ even that is analyzed in the inset of Fig.~\ref{fig_TKplaw}. Note that $T_K^{\mathrm{mean}}$ 
is the average Kondo temperature computed from the regular part of $p(T_K)$ only. 

Figure~\ref{fig_TK} shows $T_K^{\mathrm{mean}}$ for a chaotic system as a function of $\Delta/T_K^{\mathrm{bulk}}$, 
compared to the clean system, each for an even and odd electronic filling of the dot. In the bulk limit ($\Delta/T_K^{\mathrm{bulk}} \rightarrow 0$), the Kondo temperatures for even and odd $N_C$ coincide as expected, while there is a pronounced parity effect for increasing $\Delta/T_K^{\mathrm{bulk}}$, i.e., smaller systems and/or decreasing Kondo interaction strength. For the clean system, $T_K$ vanishes in the even case at the critical value $\Delta/T_K^{\mathrm{bulk}} \approx 6.28$.~\cite{mucciolo} In all other cases and for larger values of $\Delta / T_K^{\mathrm{bulk}}$, the (average) Kondo temperature scales with the level spacing $\Delta$, which becomes, in this limit, the only energy scale of the problem. 
 
The inset of Fig.~\ref{fig_TKplaw} shows the fraction $P_0$ of unscreened impurities at $T=0$ as a function of $\Delta/T_K^{\mathrm{bulk}}$. We find that it is always possible to form the Kondo singlet in the odd case (no unscreened impurities), since the chemical potential coincides with an energy level. In the even case, the Kondo effect disappears, on average, with increasing mean level spacing $\Delta$, $P_0 \rightarrow 1$.

The parity effects visible in $T_K^{\mathrm{mean}}$ are even more dramatic in the probability distribution $p\left(T_K\right)$ shown in 
Fig.~\ref{fig_TKplaw} for $\Delta/T_K^{\mathrm{bulk}}=2$ and $100$. Whereas for $\Delta/T_K^{\mathrm{bulk}}=2$ the average Kondo temperatures are almost the same, there is already a significant difference in $p\left(T_K\right)$. Most remarkable is a power law scaling, $p\left(T_K\right) \propto T_K^{-\nu}$, over several orders of magnitude in the odd case. Its existence was pointed out previously~\cite{miranda,mucciolo,cornaglia06} for a related system, the Kondo disorder model (KDM). The KDM describes a Kondo impurity in a bulk electronic bath for which disorder can be included continuously. The chaotic Kondo box that we consider here can in many respects be considered as a KDM with a non-tunable, i.e., fixed, disorder strength. Instead, another parameter can be tuned, $\Delta/T_K^{\mathrm{bulk}}$, allowing for a continuous connection between the bulk, $\Delta/W=0$, and the mesoscopic, $\Delta\sim T_K^{\mathrm{bulk}}$ regimes. 
Using Eq.~(\ref{EqTK}) we can show analytically that the power law scaling of $p\left(T_K\right)$, with the constant exponent $\nu=1/2$, is a direct consequence of the Porter-Thomas distribution of the intensities $\left|A_l\right|^2$. It persists for $\Delta \geq T_K^{\mathrm{bulk}}$ and is in agreement with Cornaglia \textit{et al.}.\cite{cornaglia06}
In the case of even occupation, a linear probability distribution, $p\left(T_K\right)\sim T_{K}$ was found for small $T_{K}$,~\cite{mucciolo} which is also consistent with our results.
For values $\Delta < T_K^{\mathrm{bulk}}$ the level structure is no longer important, so all the probability distributions are similar to a Gaussian around $T_K^{\mathrm{mean}}$.

\subsection{Magnetic susceptibility} 
\begin{figure*}
\includegraphics[width=7in]{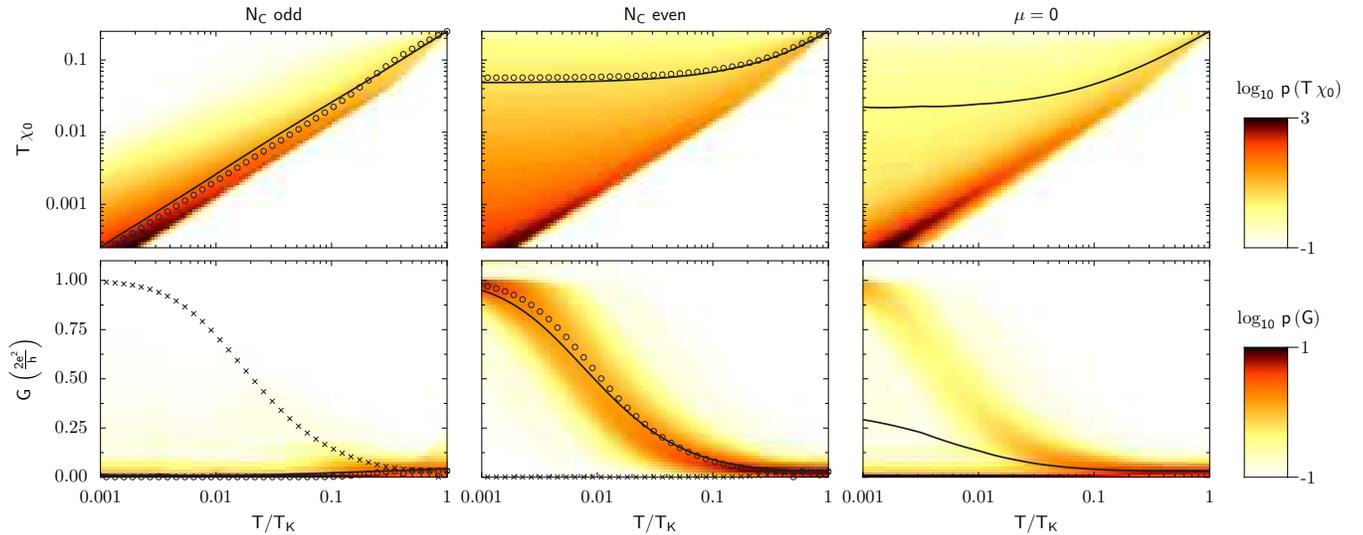}
\caption{(Color online) Distributions of the local effective moment $T\,\chi_0$, (upper panels) and the conductance $G$, (lower panels) for $N_C$ odd (left), $N_C$ even (center) and $\mu=0$ (right) as functions of $T/T_K$. Here, we use $\Delta/T_K^{\mathrm{bulk}}=2$. The clean system is symbolized by the circles ($\circ$), the solid lines mark the chaotic average values.\\
\textit{Local effective moment}: A broad distribution of values in the $N_C$ even case is contrasted with the more confined distribution for $N_C$ odd. 
This  pronounced parity effect is directly related to the formation of the Kondo singlet.\\
\textit{Conductance}: The probability distribution of the conductance at zero bias voltage shows clear parity effects.
The conductance for the non-interacting case ($J_K=T_K^{\mathrm{bulk}}=0$) is represented by the stars ($\times$), showing that the parity is switched by one, due to the Kondo impurity effective contribution. 
\label{fig_hist}
}
\end{figure*}

Within the mean-field approximation, the Kondo spin static 
susceptibility
\begin{equation}
\chi_{0}\left(T\right)\equiv \frac{1}{3}\int_{0}^{1/T} d\tau 
\langle {\bf S}\left(\tau\right)\cdot{\bf S}\left(0\right)\rangle \, ,
\end{equation}
reads
\begin{equation}
\chi_0\left(T\right) = T\sum_{i\w_n} 
G_{\mathrm{ff}}\left(i\w_n\right)G_{\mathrm{ff}}\left(-i\w_n\right) \, .
\end{equation}
In the decoupled phase, $T>T_{K}$, we recover a Curie law $\chi_{0}\left(T\right)=1/4T$, which characterizes a free spin $1/2$. In order to
describe the screening in the Kondo phase $T<T_{K}$, we analyse the 
local effective moment $T\,\chi_0\left(T\right)$, which is equal to $1/4$ for a
free spin, and vanishes for a fully screened state. Figure~\ref{fig_hist} (upper panels) shows the local effective moment 
for fixed $N_C$ as well as for $\mu=0$ at $\Delta/T_K^{\mathrm{bulk}}=2$. The clean system (circles) shows parity effects as 
expected.~\cite{italians} In the odd case, the impurity spin is fully screened at $T=0$, while in the even case it remains unscreened 
or partially screened for all temperatures. In the chaotic system the average $T\,\chi_0$ only slightly differs from the clean case, since the screening of the 
impurity spin is closely related to the number $N_C$ of dot electrons. 
More information provides the probability distribution 
$p\left(T\,\chi_0 \right)$, showing that in the odd case the 
impurity is screened at $T=0$ for all configurations in the Kondo 
regime. For $N_C$ even all values between $0$ and $1/4$ are taken
as $T\rightarrow0$. This characteristic parity dependence in $p\left(T \chi_0\right)$ should be directly 
accessible in experiments. The $\mu=0$ case contains both features: an increased probability for a screened impurity and a 
non-vanishing probability for having an unscreened impurity.

\subsection{Conductance}
Here, we consider the chaotic Kondo box connected to two leads, denoted by $\alpha=L,R$.  We study the tunneling conductance through the box in the linear response regime. The Hamiltonian of the full system is obtained from the Kondo box 
Hamiltonian~(\ref{eq:1}) by 
\begin{equation}
H \mapsto 
H+\sum_{k\sigma\alpha} \e_k \dpp{k\sigma\alpha}\dm{k\sigma\alpha} 
+ \sum_{lk\sigma\alpha} \left( t_{lk}^{\alpha} \cp{l\sigma}\dm{k\sigma\alpha} + \mathrm{h.c.}\right) \,
\end{equation}
where the second term describes the $c-$electrons of the leads and the third term the dot-lead coupling.
The leads are assumed to be ideal metals, therefore, the tunneling
couplings do not depend on the momentum $k$, $t_{lk}^{\alpha}=t_{l}^{\alpha}$. In order to mimic the chaotic nature
of the Kondo box, we also assume that the tunneling couplings are randomly distributed, with second moments 
$\langle
t_{l}^{\alpha}t_{l'}^{\alpha'}\rangle=t^{2}\delta_{ll'}$ 
(here, $\delta$ denotes the Kronecker symbol, and $t$ is a
characteristic tunneling energy). 
In the presence of a finite bias voltage 
between the leads $V$, the current through the Kondo box is \cite{meirwingreen}
\begin{equation}
J(V) = \frac{2e}{h} \int d\w \left[ f_R\left(\w\right) 
- f_L\left(\w\right) \right] \, \mathrm{Im}\left[ \Tr \left\lbrace  \mathbf{\Gamma G} \right\rbrace \right] \, ,
\end{equation}
where $f_{R/L} = f\left(\w \pm eV/2\right)$ is the Fermi function in the leads, and ${\bf G}$ is the Green's function
matrix of the dot electrons, in the presence of the leads. 
\begin{equation}
\Gamma_{ll'}^{\alpha}\equiv 2\pi \sum_k \rho\left(\e_k\right) 
t_{lk}^{\alpha}t_{kl'}^{\alpha}
\end{equation} 
is the tunneling matrix, where $\rho$ denotes the density of states of the leads. 
Hereafter, we approximate the tunneling matrix by its average value, 
$\Gamma_{ll'}^{\alpha}\approx \gamma \delta_{ll'}$, where $\gamma\sim\rho t^2$. 
A fully self-consistent mean-field treatment of the system with a
finite bias voltage would require to take into account the
renormalization of the mean-field parameters, in the presence of the
leads, similarly to the approach of Aguado \textit{et al.}.~\cite{Aguado} Here, we consider 
the linear response regime, $V\to 0$, as well as the tunneling limit, 
$\gamma\to 0$. The mean-field parameters $r$, $\lambda$ and $\mu$ are thus not modified by the leads. 
Furthermore, within the mean-field approximation, the effective Kondo box
Hamiltonian~(\ref{eq:6}) is non-interacting. Therefore, the Dyson
equation for ${\bf G}$ reads 
\begin{equation}
\mathbf{G}^{-1} = \mathbf{G}_0^{-1} - i \mathbf{\Gamma}\, ,
\end{equation} 
where the Kondo box (i.e. without the leads) $d-$electron 
Green's function ${\bf G}_{0}$ is given by Eq.~(\ref{eqGll}). 

The conductance $G = dJ/dV$ is shown in Fig.~\ref{fig_hist} (lower panels) as a function of $T/T_{K}$. 
There are clear parity effects: for $N_C$ even, the low temperature limit is $G=1$ (in units of the quantum of conductance $2e^2/h$), no matter what the underlying realization is. However, for an odd occupation number the conductance goes to $G=0$, since the Kondo singlet blocks the transport channel. 
The low temperature limit for $\mu=0$ depends on the distance between the chemical potential $\mu$ and the closest energy level.


\section{Conclusions} 
We have presented a mean-field approach to
the mesoscopic Kondo box problem that allows a very efficient computation 
of all physical quantities of interest -- $T_K$, $\chi_0$, and $G$ -- that are easily, and on an equal footing, accessed 
via the one-body Green's functions. In contrast to other methods, we are 
able (i) to reach very low temperatures $T \sim 0.001\, T_K$ within 
reasonable computation time and (ii) to calculate the probability distributions
based on a large number (at least $50\,000$) of realizations of the chaotic 
Kondo box.
Our results agree with those of other approaches when available. 
Concretely, we find deviations from the bulk system to occur in form of pronounced parity effects.
For realizations with an odd number of dot electrons, we confirm and refine the power law distribution of $p\left(T_K\right)$.
The significant parity effects in the magnetic susceptibility and the conductance provide the basis for 
a direct comparison with experiments. For example, the expected spread in the experimental values in quantum dots with a certain (even or odd) number of electrons, fixed via Coulomb blockade, will be either large or small, and is therefore already accessible by measuring few samples.
\begin{acknowledgments}
We thank Harold Baranger, Alexandre Buzdin, Stefan Kettemann, Gilles Montambaux, Eduardo Mucciolo, Jens Paaske, Achim Rosch, Pascal Simon, Grigory Tkachov, Denis Ullmo, and Matthias Vojta  for helpful and stimulating discussions. M.H. and R.B. thank the German Research Foundation (DFG) for funding within the DFG Emmy-Noether Programme. S.B. thanks the DFG for partial funding through SFB 680, SFB/TR 12, and FG 960. 
\end{acknowledgments} 


\end{document}